\documentclass[times,12pt]{article}
\pdfoutput=1
\usepackage{amsmath,amssymb,amsfonts,latexsym,amsthm,enumerate,url}
\usepackage{latex8}
\usepackage{times}
\usepackage{graphicx}
\date{}

\newcommand{\beq}[1]{\begin{equation}\label{#1}}
\newcommand{\enq}[0]{\end{equation}}

\newcommand{\remove}[1]{}

\newcommand{\comment}[1]{}

\begin{document}

\title {The Argument against Quantum Computers, the Quantum Laws of Nature, and Google's Supremacy Claims\\
{\small Laws, Rigidity and Dynamics, Proceedings of the ICA workshops 2018 \& 2019 Singapore and Birmingham}}
\author{{\large Gil Kalai\thanks{Work supported by ERC advanced grant 834735.}}
\\ {\small The Hebrew University of Jerusalem and IDC, Herzliya}\\ }
\maketitle

\begin{abstract}
My 2018 lecture at the ICA workshop in Singapore 
dealt with quantum computation as a meeting point of
the laws of computation and the laws of quantum mechanics. We described a computational
complexity argument against the feasibility
of quantum computers: we identified a very low-level complexity class of
probability distributions described by
noisy intermediate-scale quantum computers, and explained why it would allow neither
good-quality quantum error-correction nor a demonstration of ``quantum supremacy,''
(a.k.a. ``quantum advantage''), namely,
the ability of
quantum computers to make computations that are impossible or extremely hard for classical computers.
We went on to describe general predictions arising from the argument and proposed
general laws that manifest the failure of quantum computers.

In October 2019, {\it Nature} published a paper \cite {Arut+19} describing 
an experimental work that took place at Google.
The paper claims to demonstrate quantum (computational) supremacy on a 53-qubit quantum computer,
thus clearly challenging my theory. In this paper, I will explain and discuss my work in
the perspective of Google's 
claims.

\end{abstract}



\section {Introduction}

In this paper I want to present to you my theory explaining why computationally
superior quantum computing is not possible,
discuss the laws of nature that may support this theory, and describe some potential 
connections and applications.
This is a fairly ambitious task; for one, many experts do not understand my argument,
and even more
do not agree with me. On top of that,
the assertion of a paper \cite {Arut+19} published in {\it Nature} in
October 2019, declaring that ``quantum computational supremacy''
was achieved by a team from Google on a 53-qubit computer, seems to refute my argument. We will
describe and give a preliminary evaluation of Google's claims.
The story of quantum computers is related to exciting developments and problems
in physics and in the theory of computation, and my
purpose here is to tell you about it in non-technical terms (with short subsections
entitled ``under the mathematical lens'' that offer a glimpse of the mathematics and can be skipped).

\subsection {Paper outline}

Sections \ref {s:cla-comp} and \ref{s:qua-comp} introduce classical and quantum computation. Among other things, we discuss
an important heuristic concept of ``naturalness'' that is at the heart of the interface between computational complexity
and the practice of computing. Most of the paper is devoted to three related topics.
The first is my argument laid out in Section \ref {s:aaqc} of why quantum error-correction and quantum advantage are
not possible.
The second  is a description of general laws of nature
that emerge from the failure of quantum computers and quantum error-correction.
Those are described in Section \ref {s:laws} and further
connections are given in Section \ref {s:app}.
The third is a study of the Google supremacy claims. Following a description of these claims in Section \ref {s:g1}, 
we adduce in Section \ref {s:g2} reasons for thinking that the Google claims are not
reliable and discuss how to further study them. 

In Section \ref {s:nd}
we briefly discuss developments that occurred since the first version of the paper was written.
In particular, 
a 2020 paper \cite {Zho+20} in {\em Science} claimed an even greater
quantum computational advantage using a photonic system operating at room temperature.
The authors of \cite {Zho+20} claim that their device provides in 200 seconds samples that would require a
classical computer billions of years. However, it turns out that the  statistical
argument from \cite {Zho+20} is incorrect, and this is closely related to my general argument
against quantum computers. My 2014 paper with Guy Kindler \cite {KalKin14} showed a simple model demonstrating
how a classically sampled distribution may pass the same statistical tests by only reproducing small-scale
correlations of the actual theoretical distribution. 

A quick remark about terminology. The terms ``quantum supremacy''
and ``quantum advantage'' are synonyms and are both used to refer to the ability of
quantum computers to make computations that are impossible or extremely hard for classical computers.
When we refer to works by others we will adopt the terminology used in those works but, 
in view of a recent critique of the term ``quantum supremacy,'' we will otherwise use the term ``quantum advantage.''  
(``Quantum advantage'' always refers to
a computational advantage by several orders of magnitude, and
I proposed the term ``huge quantum computer advantage,'' or ``HQCA'' for short that captures the
expected magnitude of quantum advantage for some computational tasks.)
In any case, a main message of this paper
is that quantum supremacy, however referred to, is not possible.  

\medskip

{\bf Acknowledgment:} I am thankful to Yosi Atia, Ramy Brustein, Guy Kindler, Eliezer Rabinovici,
Jelmer Renema, Yosi Rinott, Tomer Shoham, and Barbara Terhal for helpful discussion.

\section {Classical computers}
\label {s:cla-comp}
\subsection {Easy and hard problems}

The central concept in the theory of computational complexity is that of an
{\it efficient algorithm} (also called ``polynomial-time algorithm'').
An efficient algorithm is an algorithm that requires a number of operations that is at most
{\it polynomial} in the size of the input.
The class of algorithmic tasks that admit efficient algorithms
is denoted by {\bf P}. For example, given a list of $n$ numbers, the task of
finding the maximal number has an efficient algorithm.

Another important algorithmic task is that
of matching. Let me elaborate a little: we are given two collections $A$ and $B$ of an equal size $n$,
and for every element $a \in A$ we are given a set $B_a \subset B$. The task at hand
is to decide whether we can find a
function $f$ from $A$ to $B$ such that 
\begin {itemize}
\item 
$f(a) \ne f(a')$ for every distinct $a$ and $a'$, 
\item  $f(a) \in B_a$ for every $a$.
\end {itemize}
Such a function is called a {\it perfect matching}. A major landmark in computer science was the
discovery by Ford and Fulkerson of an efficient algorithm for matching.

Our third algorithmic task will be the famous {\it traveling salesman problem}.
There are $n$ cities and on the road between each pair of cities $c_1$ and $c_2$ there is a
toll $T(c_1,c_2)$. A traveling salesman needs to travel between these cities,
that is, to start at city $c_1$ and then to travel through each
city exactly once, until returning to the initial city, so as to minimize the overall toll.
There is a simpler version of this problem that is called the {\it Hamiltonian cycle problem}.
For every pair of cities $c_1$ and $c_2$ we are
told in advance whether the road connecting the two is open or closed.
The challenge is  to start at city $c_1$ and then to travel through each city
exactly once, returning at the end to $c_1$ and using only open roads.
Such a route is called a Hamiltonian cycle.
A major conjecture in the theory of computational complexity is that there is
no efficient (polynomial-time) algorithm for
solving the traveling salesman problem and there is no efficient algorithm
to tell whether a Hamiltonian cycle exists. 
In fact, it is commonly believed that
an algorithm for these problems (in the most general cases) requires an
exponential number of steps and therefore
goes beyond the reach of digital computers, as the number of cities grows. 
The task of deciding whether there exists a Hamiltonian cycle constitutes
an {\bf NP-complete} problem: being in the computational class {\bf NP} means that there is 
proof that a graph $G$ has a Hamiltonian cycle that can be
verified in a polynomial number of steps. Being {\bf NP-complete}
means that any other {\bf NP}-problem can be reduced to this problem. 

\subsection {When theory meets practice: Naturalness in computer science}
\label {s:nat1}

Our main tools for the study of the complexity of algorithms are asymptotic.
For example, we make a distinction between exponential
running time and polynomial running time. When trying to gain insights into practical questions
we need to make an assumption of
{\it naturalness}, namely, that the constants involved in the asymptotic descriptions are mild.
Without such an assumption,
computational complexity insights hardly ever apply to real-life situations.
With the assumption of naturalness we do gain much insight:
if an algorithmic task can be solved (asymptotically) in a polynomial number of steps, then usually 
this suggests that the task is practically feasible. On the other hand, if a class of algorithms, or
computational devices, 
represents polynomial-time computation, then usually we cannot expect that this class of 
algorithms will practically solve intractable problems. For example,
if we are offered a device for solving the Hamiltonian cycle problem, and we can
analyze the device and realize that it represents an asymptotically polynomial-time algorithm,
then we cannot expect that this device will outperform, by a very large margin, ordinary digital computers.
Naturalness is a heuristic assertion,  
but it is a powerful one. Of course, the lower the computational
power of a class of algorithms or computing devices is
in the hierarchy of computational complexity classes, the more implausible it becomes that
such algorithms or computing devices will allow, in practice,
powerful computation.

\subsection {Randomness and computation}
One of the most important developments in the theory of computing was the realization that
the addition of an internal randomness
mechanism can enhance the performance of algorithms. 
Since the mid-1970s, 
randomized algorithms have become a central paradigm in computer science.
One of the greatest achievements was the polynomial-time randomized algorithms of
Solovay and Strassen (1977) and Rabin (1980) for
testing whether an $n$-digit integer is a prime. 
Rabin's paper stressed that the algorithm
was not only theoretically efficient but also practically excellent, and gave ``probabilistic proofs''
that certain large numbers, like $2^{300}-153$, are primes. 
This was a new kind of proof in mathematics. 

\subsection {Under the mathematical lens: Determinants and Lovasz's algorithm for perfect matching}
\label{s:match}

Let us go back to the problem of finding a perfect matching and 
consider an $n$-by-$n$ matrix $M$ where the rows correspond
to the elements of $A$, $a_1,a_2, \dots, a_n$ and
the columns correspond to the elements
of $B$, $b_1,b_2,\dots, b_n$. Now we consider variables $x_{ij}$ for every
$i,j, 1 \le i \le n, 1 \le j \le n$, and let $m_{ij}=0$ if $b_j\notin B_{a_i}$ and $m_{ij}=x_{ij}$
if $b_j\in B_{a_i}$. Lovasz's first observation was that a perfect matching exists if and only if the
determinant of $M$ (regarded as a polynomial in the variables $x_{ij}$s) is not zero.  
Lovasz's second observation was that if you create a new matrix $M'$ by
replacing $x_{ij}$ with a
random element in a large finite field, and if the
determinant of $M$ is not zero, then, with high probability, the
determinant of $M'$ is not zero either.

Here is Lovasz's algorithm: given the data, we build at random the matrix $M'$ and
check whether its determinant equals zero and repeat this process $k$ times.
If we get a non-zero answer once, we know that there is a perfect matching; if we always get 
zero, we know with high probability that a perfect matching does not exist.

We need one additional ingredient that goes back to Gauss: when the entries are concrete numbers, 
there is a polynomial-time algorithm for
computing determinants. This is based on Gauss's elimination method, and can be considered
as one of the miracles of our world.

\section {Quantum computers}
\label {s:qua-comp}
\subsection {Huge computational advantage: Factoring and sampling}

Quantum computers are hypothetical physical devices that allow
the performance of certain computations well beyond the ability of classical computers,
in a polynomial number of steps in the input size.  
The basic memory unit of a quantum computer is called a {\it qubit} and the 
basic computational step on one or two such qubits 
is performed by {\it gates} (further details are given below). 
Shor's famous algorithm shows that quantum computers can
factor $n$-digit integers efficiently, in roughly $n^2$ steps! (The best known classical
algorithms are exponential in $n^{1/3}$.) This ability for efficient factoring allows quantum computers to break
the majority of current cryptosystems.

A {\it sampling task} is one where the computer (either quantum or classical) produces samples
from a certain probability distribution $D$. In the main examples of this paper
each sample is a 0-1 vector of length $n$, where $D$ is a probability distribution on such vectors.
Quantum algorithms allow sampling from probability distributions well beyond the capabilities of classical
computers (with random bits). Shor's algorithm exploits the ability to
sample efficiently on a quantum computer a probability distribution based on   
the Fourier coefficients of a function.

\subsection {Noisy quantum computing}

Quantum systems are inherently noisy: we cannot accurately
control them, nor can we accurately describe them. In fact, every interaction of a quantum
system with the outside world amounts to noise.
A noisy quantum computer has the property that every computational step 
(applying a gate, measuring a qubit) makes an error with a certain small probability $t$. 
(These errors are described more specifically in Section \ref {s:g1}, whereas in Section \ref {s:laws-ml}
we get a glimpse of the mathematics of noise in quantum systems.)
The threshold theorem \cite {AhaBen97,Kit97,KLZ98} asserts that
if the rate of errors $t$ is small enough (and if a few additional
assumptions are made), then a noisy quantum circuit can simulate
noiseless quantum circuits. To implement such a simulation we need certain building blocks 
called {\it quantum error-correcting codes,}
where a collection of 100--5000 quantum qubits (or more) can be ``programmed'' 
to represent a single stable ``logical'' qubit.

\subsection {NISQ computers}

{\it Noisy intermediate-scale quantum (NISQ)} computers, are quantum computers
where the number of qubits is in the tens or
at most in the hundreds.
Over the past decade researchers have conjectured \cite {AarArk13} that the huge
computational advantage of sampling with quantum computers 
can be realized by NISQ computers that only approximate the 
target probability distribution. These researchers have predicted that 
quantum computational advantage 
(for sampling tasks)
could be achieved for NISQ computers without using quantum error-correction.
NISQ computers are also 
crucial to the task of creating good-quality quantum error-correcting codes. An important 
feature of NISQ systems -- especially for the tasks of achieving quantum advantage
and quantum error-correction -- is the fact that a single error
in the computation sequence has a devastating effect on the outcome. In the NISQ regime,
the engineering task is to
keep the computation error-free. We shall refer to the probability
that not even a single error occurs as the {\it fidelity}. 

Many companies and research groups worldwide are implementing quantum computations via
NISQ computers (as well as by other means). There are several different approaches to realizing individual
qubits and gates, and each of the main
approaches is marked by different variations. Realizing quantum circuits by
superconducting qubits is a leading approach, whereas 
trapped-ion qubits, photonic qubits, topological qubits, and others are considered notable alternatives.

\subsection {Under the mathematical lens: The mathematical model of quantum computers}

\subsubsection {Quantum computers (circuits)}

\begin {itemize}

\item
A {\it qubit} is a piece of quantum memory.
The state of a qubit is a unit vector in a two-dimensional vector space over the complex numbers $H = \mathbb C^2$.
The memory of a quantum computer (quantum circuit) consists of $n$ qubits and the
state of the computer is a unit vector
in the $2^n$-dimensional Hilbert space, i.e.,  $\mathbb (C^2)^{\otimes n}$.

\item
  A {\it quantum gate} is a unitary transformation.  We can put one or two qubits through gates, which
  represent unitary transformations, that act on the
corresponding two- or four-dimensional Hilbert spaces.
There is a small list of gates that are sufficient for the full power of quantum computing.
\item
{\it Measurement} of the state of $k$ qubits leads to a probability distribution on 0-1
vectors of length $k$.

\item
A {\it quantum circuit} is composed of a collection of gates acting successively on $n$ qubits.
To describe an efficient (or polynomial-time)
quantum algorithm, we assume
that the number of gates is at most polynomial in $n$. (We
also assume that the sequence of gates can be produced efficiently by a classical algorithm.)
\end {itemize}

\subsubsection {Superposition and entanglement}

The state of a single qubit is a {\em superposition}
of basis vectors of the form
$a  \left|0\right\rangle +b  \left|1\right\rangle $,
where $a,b$ are complex and $|a|^2+|b|^2=1$.
The complex coefficients $a$ and $b$ are called {\it amplitudes}. A  measurement of
a qubit in state $a  \left|0\right\rangle +b  \left|1\right\rangle $ will lead to a random bit of 0
with probability $|a|^2$ and 1 with probability $|b|^2$. This rule for moving from complex amplitudes to
probabilities is referred to as the ``Born rule.''

Two qubits are represented by a tensor product $H \otimes H$ and we denote  $  \left|00\right\rangle =
\left|0\right\rangle \otimes \left|0\right\rangle$.
The 
{\it cat state}
${\frac{1}{\sqrt 2}}\left|00\right\rangle +{\frac{1}{\sqrt 2}} \left|11\right\rangle$
can be regarded as a quantum analog, called {\it entanglement}, of correlated
coin tosses that yield two heads
with probability 1/2, and two tails with probability 1/2. The cat state is
the simplest example of entanglement, and the strongest form of entanglement between two qubits.

\section {The argument against quantum computers}
\label {s:aaqc}

\subsection {My argument against quantum advantage and quantum error-correction}

Here, in brief, is my argument against quantum computers. For more details see \cite {Kal20,Kal18}.

\begin {quotation}
\noindent

{\bf (A)} From the perspective of computational complexity theory,
noisy intermediate-scale quantum (NISQ) circuits are low-level classical computational devices.
\end {quotation}

\begin {quotation}

\noindent

{\bf (B)} Therefore, by naturalness, NISQ systems do not support quantum advantage.
In other words, the rate of noise cannot be reduced to the level allowing quantum advantage.
\end {quotation}

\begin {quotation}
\noindent

{\bf (C)}
Achieving good-quality quantum error-correction requires an even lower noise
rate than the one required for achieving quantum advantage.
\end {quotation}

\begin {quotation}
{\bf (D)}
Therefore, NISQ systems do not support quantum error-correction.
\end {quotation}

\begin {quotation}
{\bf (E)}
Hence, large-scale quantum computing based on quantum error-correction is beyond reach.
\end {quotation}

\subsection {Four thresholds}

To put the above argument a little differently, we can consider four crucial thresholds of noise,
$\alpha, \beta, \gamma, \delta $:\footnote{$\alpha,\beta,\gamma$, and $\delta$ are not universal constants; they
  depend (moderately) on a specific implementation. Our argument asserts that inequality (\ref{e:ineq}) holds universally.}   
\begin {itemize}
\item
$\alpha$ is the rate of noise required for 
universal quantum computing, 
\item
$\beta$ is the rate of noise required for good-quality quantum error-correction, 
\item
$\gamma$ is the rate of
noise required for quantum advantage, 
and 
\item
$\delta$ is the rate of noise that can realistically be achieved.
\end {itemize}

Since universal quantum computing requires very
good-quality quantum error-correcting codes, we get that $\alpha < \beta$.
At the center of my analysis is a computational complexity argument stating that $\gamma< \delta$,
and I also rely on the argument that 
$\beta < \gamma$, which is in wide agreement. Given these inequalities, we get that 
\begin {equation}
  \label {e:ineq}
\alpha < \beta < \gamma < \delta.
\end {equation}

We note that it is a strong intuition of many researchers that
with sufficient engineering efforts, $\delta$ can be reduced
as close to zero as we want. My argument implies that this belief is incorrect.

\subsection {Four facts that strengthen the argument}
\label {s:4f}
There are four facts that strengthen this argument against quantum computers. 
\begin {itemize}
\item[1.]
The first is that NISQ circuits are
very, very low-level classical computational devices.  
\item[2.]
The second is that while our argument asserts that  the level of noise 
that can realistically be achieved will be above the level of noise allowing
the demonstration of quantum advantage, 
there is yet another, related argument asserting that
when we consider $n$-qubit circuits, then for a wide range of
lower levels of noise, the outcomes will be {\it chaotic}: no robust
probability distributions will be possible as the output. 
\item[3.]
The third fact is that there are also direct reasons why probability distributions 
supported by quantum error-correcting codes (like the popular ``surface codes'') are
not supported by the very low-level  
computational complexity class of NISQ circuits. 
\item[4.]
The fourth fact is that while quantum error-correction requires achieving very high fidelity 
for tens or hundreds of qubits, it has been realized in recent years (and this forms the very
basis for Google's experiment) 
that quantum advantage can be demonstrated even with low fidelity.
\end {itemize}

The first and second items in the list are the most important, and I would therefore
like to say a little more about them.
(The reader is referred to our next mathematical Section \ref {s:aaqc-ml} and
to \cite {Kal20,Kal18} for more details.)

The computational complexity class describing NISQ circuits is {\bf LDP} (low-degree
polynomial) and this class
is contained in the familiar class of distributions
that can be approximated by bounded-depth (classical) computation. 
 
Let me phrase the second point a little differently. The threshold for realistic noise $\delta$ cannot 
be pushed down to allow quantum advantage; but more than that is true:  there 
is a large range of error rates below $\delta$,
where even if you could reduce the error rate to these levels,  
the resulting probability distribution would be chaotic and would largely depend on the fine parameters 
of the noise itself.

\subsection {Under the mathematical lens: Noise stability and sensitivity and Fourier--Walsh expansion.}
\label {s:aaqc-ml}

The first assertion in my argument is related to a mathematical theory of 
noise stability and noise sensitivity that goes back to
Benjamini, Kalai, and Schramm (1999) \cite {BKS99} (and can be traced back to \cite {KKL88}).
In my lecture in Singapore
I described this theory in the context of voting methods. How likely is
it that the outcome of an election will be
reversed because of noise in counting the votes?

Let $\Omega_n$ be the set of 0-1 vectors of length $n$.
We start with a real function $f(x_1,x_2,\dots, x_n)$, and for a real
number $t$, we define the noise version of $f$ as

\begin {equation}
\label {e:noise-basic}
N_t(f)(x)=\sum_{y \in \Omega_n}f(x+y)t^{|y|}(1-t)^{n-|y|}.
\end {equation}

Here $y=(y_1,y_2,\dots,y_n)$ is also a 0-1 vector and $y_i=1$ indicates ``error in the $i$th coordinate.''
The sum $x+y$ should be considered as a
sum modulo 2: $x_i+0=x_i$ and $x_i + 1= 1-x_i$, and $|y|=x_1+x_2+\cdots+x_n$.

It turns out (\cite{BKS99}) that the behavior of noise for functions on $\Omega_n$  is closely related to
the Fourier--Walsh expansion of the function. Here is a quick description.
Recall that for $S\subset [n]=\{1,2,\dots,n\}$,
the Walsh function $W_S$ is defined
as
\begin {equation}
  W_{\emptyset}=1~~~ {\rm and}~~~ W_S(x_1,x_2,\dots,x_n)=\prod_{i \in S}(1-2x_i).
\end {equation}
If the Fourier--Walsh expansion of $f$ is
\begin {equation}
\label {e:fourier-Walsh}
f=\sum_{S \subset [n]} \hat f(S)W_S,
\end {equation}
\noindent
then
\begin {equation}
\label {e:noise-fourier}
N_t(f)=\sum_{S \subset [n]} \hat f(S)(1-2t)^{|S|}W_S.
\end {equation}

\begin{figure}
\centering
\includegraphics[scale=0.12]{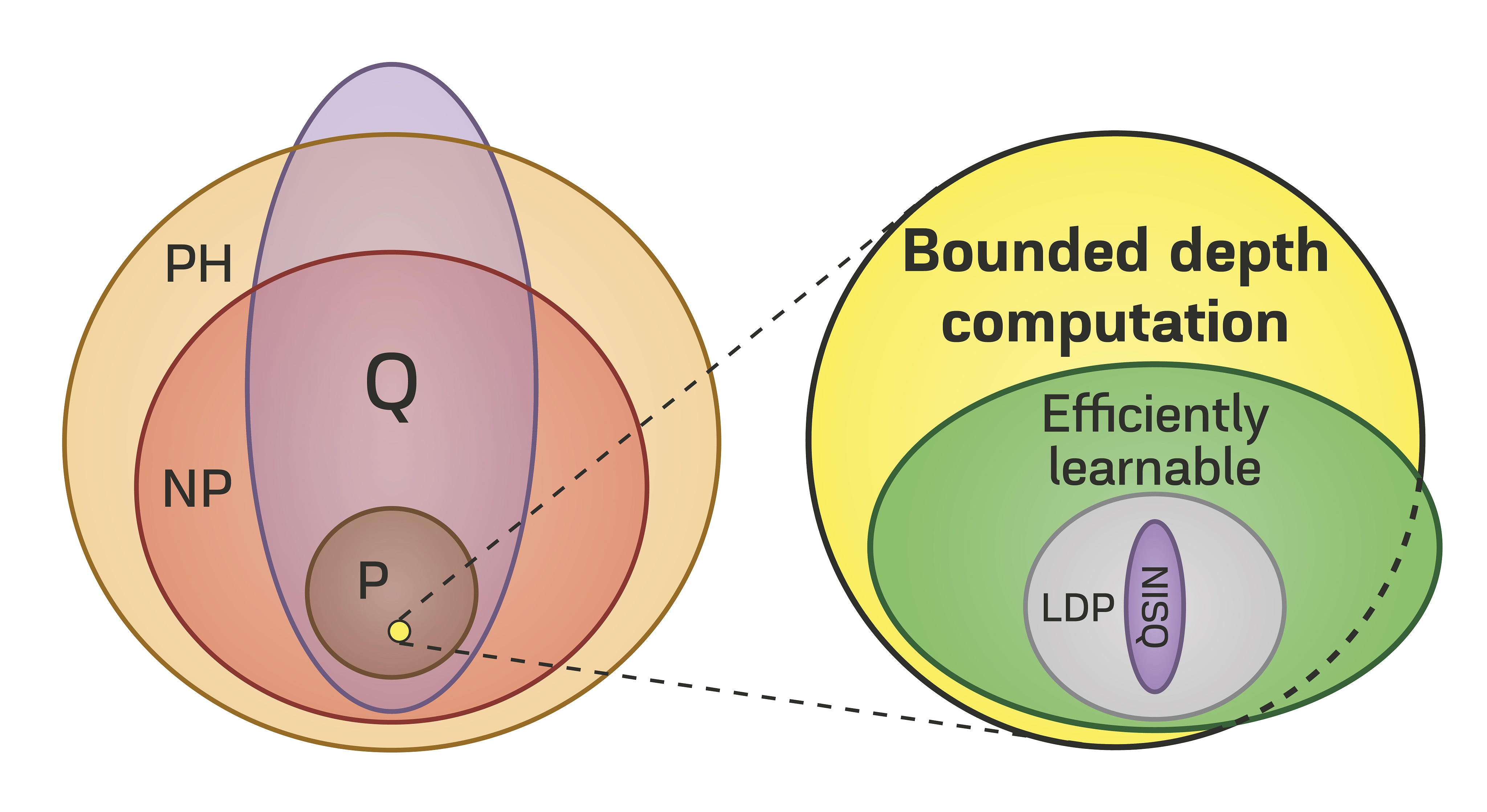}
\caption{Probability distributions described by NISQ systems represent
a low-level computational class {\bf LDP}.
Approximate sampling for {\bf LDP}-distributions belongs to the class of
bounded depth (classical) computation  
and is also efficiently learnable.}
  \label{fig:1}
\end{figure}

If the value of $f$ is always 0 and 1 we call $f$ a Boolean function, then
we can regard $f$ as a voting rule 
for a two-candidate election. A deep finding from \cite {BKS99} is that for a wide class of voting rules, 
only those voting rules that are close enough to the ``majority'' voting
rule (or a weighted version of the majority rule) 
are noise-stable. We note that the majority voting rule is related to the very basic methods for 
achieving robust classical information and computation.

We can now describe 
the ``very low-level'' computational complexity class {\bf LDP}
of probability distributions described by NISQ systems. The class {\bf LDP} consists of
probability distributions that
can be approximated by polynomials of bounded degree. 
Indeed, when $t>0$ is fixed and we apply the noise $N_t$ (defined by Equation (\ref{e:noise-basic}))
to an arbitrary probability distribution $D$, the resulting distribution $N_t(D)$ can be
well approximated by polynomials of bounded degree (roughly $1/t$). (This easily follows
from Equation (\ref{e:noise-fourier}).)
Distributions that can be (approximately) described by bounded-degree polynomials
can also be approximately described by {\it bounded-depth (classical) circuits}. (Bounded depth circuits
define a well-known low-level complexity class ${\bf AC^0}$.) Approximate sampling for {\bf LDP}-distributions
is also {\it efficiently learnable,} which also describes a low level complexity class for approximate sampling. (See Figure
\ref {fig:1}).

When $D$ is a probability distribution proposed for ``quantum advantage'' (or arising from
quantum error-correcting codes),
then, even when the level of noise is subconstant but (well) above $1/n$, the
correlation between the two distributions $D$ and $N_t(D)$ tends to zero. This suggests
that for realistic forms of noise the noisy probability distribution will strongly
depend on fine parameters of the noise itself, leading to a ``chaotic'' behavior.

We note that the analysis of noise sensitivity of NISQ systems was initially carried out 
on another model called ``boson sampling'' in \cite {KalKin14}. 
For further discussion of boson sampling see \cite{AarArk13,TroTis96,Kal20,Kal16,Kal18}, and Section \ref {s:nd}.

\section {The laws}
\label {s:laws}

Without further ado let us now move to the laws of physics
that emerge from the failure of quantum computers.

\subsection* {Law 1: Time-dependent 
quantum evolutions are inherently noisy.}

\subsection* {Law 2: Probability distributions described
by low-entropy states are noise-stable and can be expressed by low-degree polynomials.}

\subsection* {Law 3: Entanglement is accompanied by correlated errors.}

\subsection* {Law 4: Quantum noise accumulates.}

We emphasize that these four laws are compatible with quantum mechanics.
The laws proposed in this section 
are not part of the argument
for why quantum error-correction is not possible, but largely rely on taking
this argument for granted.

\subsection { The first law: Time-dependent 
quantum evolutions are inherently noisy.}

Time dependence in a quantum evolution amounts to an interaction with the environment and 
the first law asserts that there is no way around the noise -- not for a single qubit and not for more
involved quantum evolutions. 
In Section \ref {s:laws-ml2} below we briefly suggest how to put the first law 
on formal grounds.

\subsection {The second law: Probability distributions described
by low-entropy states are noise-stable and can be approximated by low-degree polynomials.}

The second law  extends our assertions regarding NISQ circuits beyond the NISQ regime. 
The noise causes the 
high-degree terms, in a certain Fourier-like expansion of the probability distribution, 
to be reduced exponentially with the degree. Low-entropy states, 
for which the effect of the noise is small, 
have probability distributions expressed by low-degree Fourier
terms.\footnote{For the definition of entropy see Section \ref {s:laws-ml1}.
Meanwhile, we can think of ``low entropy'' as a synonym for ``high fidelity.''}  
Such noise-stable states represent the very low-level computational complexity 
class,  {\bf LDP}, namely, the class of probability distributions 
that can be approximated by low-degree polynomials. 
Our second law applies to quantum evolutions in nature that can be described by quantum circuits, 
and it is a plausible assumption that this applies
universally (under some caveats; see Section \ref {s:ste}). 
We can expect that the specific ``Fourier-like expansion'' 
will be different for different physical settings
but that the same computational class {\bf LDP} will apply in general.

\subsection {The third law: Entanglement is accompanied by correlated errors.}
\label {s:law3}

The third law  asserts that the errors for the two qubits of a cat state necessarily 
have a large positive correlation. Here also we extend well-accepted insights of NISQ systems
into general quantum systems. 
Correlated errors is an observed 
and accepted phenomenon for gated qubits and, without
quantum error-correction, it extends and applies to all pairs of entangled qubits. 
An important consequence of the third law is that 
complicated quantum states and evolutions lead to {\it error synchronization}, namely, to a substantial 
probability that a large number of qubits, far beyond the average rate of noise, are hit by
noise.

We emphasize that the third law is not based on a new way to model noisy quantum circuits, but rather
is derived from ordinary models under the assumption that $\beta < \delta$, namely, that the
error rate cannot be reduced to the level that enables quantum error-correction. It would be interesting to
test the quantitative aspects of the law both by simulation and by experiments.
See also Section \ref {s:n-cor}.
We note that this law is related to our proposed modeling
in Section \ref{s:g2-ml} (Equation (\ref{e:nqc2})) but 
is not related to correlations in the computation of the fidelity via Formula (77), that we
discuss, in the context of Google's experiment, in Sections \ref {s:g1} and \ref {s:g2}.

\subsection {The fourth law: Quantum noise accumulates.}

The fourth law expresses the fact that without noise cancellation via 
quantum fault-tolerance, quantum noise must accumulate. 
In Section \ref {s:laws-ml2} we briefly suggest how to put the fourth law 
on formal grounds.

\subsection {Under the mathematical lens: Noise, time, and non-commutativity} 

\label {s:laws-ml}

\subsubsection {Noise, mixed states, density matrices, and entropy}

\label {s:laws-ml1}

In quantum physics, states and 
their evolutions (the way they change over time) 
are governed by the Schr\"odinger equation. 
A solution of the 
Schr\"odinger equation can be described 
as a unitary process on a Hilbert space, and the states (which are called ``pure states'') 
are simply unit vectors in this Hilbert space.  
Quantum computers, as described above, 
form a large class of such quantum evolutions, and it is even a common view
that all quantum processes in nature (or at least all ``local'' quantum processes)
can be described efficiently by quantum computers. 
When you add noise to the picture you encounter more general types of states (called ``mixed states'') 
that can be described (not in a unique way) as a classical probability distribution 
of pure quantum states.\footnote{An alternative description of noisy states and evolutions 
can be given in terms of a larger Hilbert space $H' \supset H$, and a unitary process on $H'$.}  
Mathematically speaking, if $\rho$ is a pure state and hence a (row) unit vector 
in (say) an $N$-dimensional space, we represent $\rho$ by the matrix $\rho^{tr}\rho$. 
(This matrix is the outer product of $\phi$ with itself; in the quantum ``bra-ket'' notation we write it as
$ \left| \rho \rangle \langle \rho \right| $.)
A convex combination of such matrices 
represents a general mixed state and this representation is referred to as the density matrix 
representation.\footnote{Quantum evolutions on density matrices are described
  by ``quantum operations.'' We will 
  not discuss them  here, but merely mention that
  their study was the starting point of central areas in mathematics.}  
The von Neumann entropy $S(\rho)$ of a state $\rho$ (in terms of the density matrix description) 
is defined by $S(\rho)=-tr (\rho \log \rho)$. (Here we refer to 
logarithm as a function on matrices and logarithm is taken to the base 2.) The entropy is always non-negative 
and, for a state $\rho$, $S(\rho)=0$ if and only if $\rho$ is a pure state.

\subsubsection {Commutativity, time, and time-smoothing}

\label {s:laws-ml2}

I will now briefly describe some mathematical ideas required for putting  
the first and fourth laws on more formal grounds. 
One obstacle we face when trying to mathematically express the claim that
time-dependent evolutions are noisy is that 
the parameterization of time we start with is arbitrary. We need 
to consider a canonical parameterization of time. 
Now, you may recall that two operators $U$ and $W$ (or matrices) are commutative if $UW=WU$. For two
operators $U$ and $W$ that do not commute (namely, $UW \ne WU$)
a non-commutativity measure refers to a quantitative way to measure by how much $U$ and $W$
fail to commute.

\subsection* {The first law (reformulated):
Noise in a certain time interval is bounded from below by a non-commutativity measure of the involved unitary operators.}

Furthermore, such a non-commutativity measure can be regarded as 
an intrinsic parameterization of time for a quantum evolution.

The first law asserts that when you look at a quantum computer that in a certain time
interval executes a sequence of unitary operators $U_1,U_2,..., U_s$, then
the amount of noise at that time interval is bounded from below by a non-commutativity measure of those unitary operators. 
If you start with a single qubit and apply a random sequence of 1-qubit gates you
recover the assertion that the quality of a qubit has an absolute positive lower bound. 
Time dependence allows us to formulate a general law for lower bounds on the amount of noise and to put the intuition that
quantum systems are inherently noisy on formal grounds. We note that the first law does not imply that
every time-independent quantum evolution can be realized without noise.

We end the section with a brief discussion of the fourth law.
The fourth law asserts 
that quantum noise must accumulate and that noise cancellation via quantum fault-tolerance
is not possible. 
To express this idea mathematically we model ``noise accumulation'' by considering a subclass of all noisy quantum 
evolutions where the noise is given by a certain time-smoothing operation.

\subsection* {The fourth law (reformulated): 
Noisy quantum evolutions are subject to convoluted time-smoothed noise.}

Convoluted time-smoothing is a certain mathematical operation that averages out the error over time. 
(For the definition see \cite {Kal18}[Sec. 4.6.2], \cite {Kal16}.) 
The crucial property is that the ``convoluted time-smoothing'' can 
be applied to every noisy quantum evolution, but not every noisy quantum evolution 
is obtained by such smoothing. We thus end up with a subclass 
of all noisy quantum evolutions, which is suggested as a class of  
evolutions where quantum noise necessarily accumulates.
We face the difficulty that for general quantum evolutions
time parameterization is arbitrary and, here too, we need to take
the parameterization of time for the smoothing to be intrinsic.

\section {The Google supremacy claims}
\label {s:g1}

\subsection {The experiment}

The Google experiment is based on the building of a quantum computer (circuit) with $n$ qubits 
that performs $m$ rounds of computation.
The computation is carried out by a 1-qubit and 2-qubit gates. 
At the end of the computation the qubits are measured, leading 
to a probability distribution on 0-1 vectors of length $n$. 
For the ultimate experiment ($n=53$, $m=20$, 1113 1-qubit gates, 530 2-qubit gates) 
the Google team produced a sample of a few million 0-1 vectors of length 53.

The specific circuit $C$ used for the computation is a random circuit. For every experiment, 
the specific gates are chosen, once and for all, at random (by a classical computer). Without noise
the quantum computer will produce samples from a certain probability distribution 
$D_C$ that depends on the specific circuit $C$.  Google's quantum computers (like any other
quantum computers currently available) 
are ``noisy,'' so what the computer is
actually producing are not samples from $D_C$ but rather a noisy version that
can roughly be described as follows:
a fraction $F$ of the samples are from $D_C$
and a fraction $(1-F)$ of the samples are from a uniform distribution. $F$
is referred to as the {\it fidelity}.

\subsection {The Google supremacy claims}

The paper made two crucial claims regarding the ultimate 53-qubit samples.

\begin {itemize}
\item[A)] The fidelity $F$ of their sample is above $1/1000$.

\item [B)] Producing a sample with similar fidelity 
would require 10,000 years on a supercomputer.
\end {itemize}

\subsection {Google's argument}

As it was only possible to give indirect evidence for both these claims, we shall now describe the
logic of Google's quantum supremacy argument.

For claim A) regarding the value of $F$, the paper describes a statistical estimator for $F$ 
and the argument  relies on a bold extrapolation argument that has two ingredients. 
One ingredient is a few hundred experiments in the classically tractable regime: the 
regime where the  probability distribution $D_C$ can be
computed by a classical computer and the performance of the 
quantum computer  can be tested directly. The other
ingredient is a theoretical formula for computing the fidelity.
According to the paper, the fidelity of entire circuits closely agrees with the prediction of the simple
mathematical formula (Formula (77) in \cite {Arut+19S}; Equation (\ref{e:77}) below) 
with a deviation below 10--20 percent. 
There are around 200 reported experiments in the
classically tractable regime including ones carried out on simplified
circuits (which are easier to simulate on classical computers). These experiments
support the claim that the
prediction given by Formula (77) for the fidelity is indeed very robust and applies
to the 53-qubit circuit in the supremacy regime.
We note that the samples for the 53-qubit experiment demonstrating ``supremacy'' 
are archived, but that it is not possible to test them 
in any direct way. 

For claim B) regarding the classical difficulty, the Google team
mainly relies on extrapolation from the running time of a
specific algorithm they use. They also rely on the computational complexity support for the assertion
that the task at hand is asymptotically difficult. (It is also to be noted 
that using conjectured asymptotic behavior for 
insights into the behavior in the small and intermediate scales relies
on a naturalness assumption.)   

\subsection {Estimating the fidelity}

\subsubsection *{Google's statistic $F_{XEB}$.}

Once the quantum computer produces $m$
samples $x_1,x_2,\dots, x_m$, the following estimator for the fidelity is computed:
\begin {equation}
F_{XEB}= 2^n\frac {1}{m}\sum_{i=1}^m D_C(x_i) -1.
\end {equation}

\subsubsection *{Google's a priori fidelity prediction}
\label {s:77}

The Google argument relies crucially on the following simple formula (Formula (77) in \cite {Arut+19S}) 
for estimating the fidelity $F$ of their experiments:

\begin {equation}
\label {e:77}
~F~=~ \prod_{g \in {\cal G}_1} (1-e_g) \prod_{g \in {\cal G}_2} (1-e_g) \prod_{q \in {\cal Q}} (1-e_q).
\end {equation}

Here ${\cal G}_1$ is the set of 1-gates (gates operating on a single qubit), ${\cal G}_2$ is 
the set of 2-gates (gates operating on two qubits), and
${\cal Q}$ is the set of qubits. For a gate $g$, the term $e_g$ in the 
formula refers to the fidelity (probability of an error) of the individual gate $g$. 
For a qubit $q$, $e_q$ is the probability of a read-out error when we measure the qubit $q$.
If we replace the detailed individual values for the fidelities by their average
value we get a further simplification:

\begin {equation}
\label {e:77s}
F'= (1-0.0016)^{|{\cal G}_1|} (1-0.0062)^{|{\cal G}_2|} (1-0.038)^n .
\end {equation}

The rationale for Formula (77) (Equation (\ref {e:77})) 
is simple: as long as there are no errors in the performance of all 
the gates and all the measurements of the qubits, then we get a sample from the
correct distribution. A single 
error in one of these components leads to an irrelevant sample.
The Google paper reports that for a large number of experiments the actual fidelity 
estimated by Formula (77) (Equation (\ref {e:77}))  
agrees with the statistical estimator for the fidelity up to 10\%--20\% percent.
We can expect 
that the value of $F'$ will be a few percentage points higher than
that\footnote{An even better approximation is $(1-0.0093)^{|{\cal G}_2|} (1-0.038)^n$.} of $F$.
For the 
circuits
used by 
Google, when the number of qubits is $n$ and the number of 
layers is $m$ ($m$ is an even integer), $|{\cal G}_1| = n(m+1)$ 
and $|{\cal G}_2| \le nm/2$.

\subsubsection *{Google's statistical philosophy}

A basic statistical idea in the Google paper (\cite {Arut+19S}, [Sect. IV]) is the following:

\begin {quotation}
  
Crucially, XEB does not require the reconstruction of experimental output probabilities, which
would need an exponential number of measurements for increasing number of qubits.
Rather, we use numerical simulations to calculate the likelihood of a set of bitstrings obtained in an
experiment according to the ideal expected probabilities.

\end {quotation}

\subsection {Under the mathematical lens: 
The Porter--Thomas probability distributions, 
Archimedes, and size-biased distributions}
\label{s:g1-ml}
\subsubsection *{What does a ``random'' probability distribution look like?}

Let $X$ be a set and our task will be to describe a ``random'' probability distribution $D$ on $X$. 
Consider another real probability distribution $Z$
where $Z$ is a positive real number and $\mathbb E(Z)=1$.
Now, to $x \in X$ we assign a probability $z(x)/|X|$ drawn at
random from $Z$. (To make sure that those are indeed 
probabilities you need to normalize
 $\sum_{x \in X} z(x)$ to $1$.) 
This construction was made in a nuclear physics paper by Porter and
Thomas (1956) \cite {PorTho56} for the case where $Z$ 
is an $\chi^2$-distribution.  
The general construction was made in a statistics paper by 
Kingman (1975) \cite {Kin75}.

In the case of Google's experiment, $X=\Omega_n$ (the set of all 0-1 vectors of length $n$) and 
$Z$ is the exponential distribution with density function $e^{-z}$. Also, $D$ is not really random:
it is a pseudorandom distribution with properties very similar to those of a truly
random distribution. Here, by pseudorandom we mean a value, drawn
by a computer program, that behaves ``like'' a random value.
The twist here is that the computer program is a quantum computer program.
The assumption behind the quantum advantage claims is that computing
this pseudorandom distribution is a very
hard problem for a classical computer, yet sampling from this distribution
can be easily carried out by a quantum computer. 

\subsubsection *{The exponential distribution, Archimedes, and moment maps}

The state of a quantum computer that performs a random sequence of gates is similar to a random unit vector
in the Hilbert space described by the computer. Now, when you consider
a random unit vector in a high-dimensional complex vector space,
the distributions of the real and complex parts of each coordinate are close to
Gaussian and, therefore, the distribution of their sum of squares
is exponential. Indeed, recall that, in general,
the sum of squares of $k$ statistically independent
Gaussians is $\chi_k$, the $\chi$-square distribution with
$k$ degrees of freedom, and for $k=2$ this is the exponential
distribution. (The statistical independence condition
approximately holds for random unit vectors in high dimensions.)

There is a further interesting mathematical story related to 
why the probabilities $D_C(x)$ behave according to a Porter--Thomas distribution based on an exponential
distribution $Z$. The space $\Delta$ of all probability distributions
on $\Omega_n$ is a simplex of dimension $2^n-1$.
Now, consider a point, drawn at random from a unit sphere in a complex
space of dimension $2^d$. 
When we replace ``amplitudes'' (complex coefficients) by
the associated real probabilities, 
we obtain (precisely, on the nose) a random probability distribution, namely,
a point from this simplex $\Delta$ drawn uniformly at random.
As pointed out by Greg Kuperberg \cite {Kup19}, the connection between the complex amplitudes and the 
probability distribution is related to 
a theorem of Archimedes (c. 287  -- c. 212 BC), whereby a natural projection from the
unit sphere to a circumscribing vertical
cylinder preserves area. (It is also related to the ``moment map'' in modern symplectic geometry.)

\subsubsection* {Statistics: Size-biased distributions}

Let us suppose that you want to estimate the distribution $D$ of the number of people in apartments. 
You sample random people on the street and ask each one
how many people share his apartment with him.
The distribution, $E$, of answers will not be identical to $D$:  
a quick way to see this is based on the fact that people you meet on the
street are not from empty apartments. 
We face a similar situation when we let the quantum computer
sample $x \in \Omega_n$ (this is an analog to the random person
we meet on the street) and then compute $D_C(x)$
(this is
an analog to asking about how many people share his apartment). The resulting 
size-biased distribution is given by $ \Gamma = xe^{-x}$,
and constitutes the basis for the statistical estimator $F_{XEB}$ 
for the fidelity $F$. For more on size bias see \cite {AGK19,RSK20}.

\subsubsection *{Google's statistics $F_{XEB}$.}

Recall that once the quantum computer produces $m$
samples $x_1,x_2,\dots, x_m$, the following statistic is computed:
$$F_{XEB}= 2^n\frac {1}{m}\sum_{i=1}^m D_C(x_i)-1.$$
The expected value of $2^n D_C(x)$ when $x$ is drawn uniformly at random is 
$$\int_0^{\infty}xe^{-x}dx=1,$$
while the expected value of $D_C(x)$ when $x$ is drawn from the distribution $D_c$ itself is 
$$\int_0^{\infty}x^2e^{-x}dx=2.$$
It follows that when $x$ is drawn from the distribution $FD_C + (1-F)U$,
the expected value of $2^nD_c(x)$ is $1+F$ 
and, therefore, $F_{XEB}$ is an unbiased estimator for the fidelity $F$.

\section {Preliminary assessment of the Google claims}
\label {s:g2}

The Google experiment represents a very large leap forward with regard to
several aspects of the human ability to control
noisy quantum systems. Accepting the Google claims requires a very careful evaluation
of the experiments and, of course, successful replications as well.
The burden of producing detailed documentation of the experiments and 
carefully examining the experimental data and that of replications
lies primarily with the Google team itself and, naturally,
also with the scientific community as a whole.

In my view, there are compelling reasons
to doubt the correctness of the Google supremacy claims.
Specifically, I find the evidence for the main supremacy
claim A) concerning the 53-qubit samples too weak to be convincing. 

Furthermore, in my opinion, there are compelling reasons to question
the crucial claims regarding perfect proximity between predictions based
on the 1- and 2-qubit fidelity and the circuit fidelity. 
Some of the outcomes reported in the paper appear to be ``too good to be true''; that is,
the experimental outcomes are unreasonably close to the expectations of the experimentalists.
In this section we shall focus on the main example of this type.

It is to be noted that there are also several works that challenge Google's claim B)
regarding the complexity of their sampling task on a classical computer. 
A team from IBM \cite {PGN+19} demonstrated a way of improving the running
time by 6 orders of magnitude.  Another group \cite {ZSW20} 
demonstrated an improvement all the way to within 1--2 orders of magnitude
above the quantum running time 
for a related (albeit easier) sampling problem. Yet another group \cite {Hua+20} proposed a tensor network-based classical
simulation algorithm for Google's circuit. (See also Section \ref {s:nd}.)

\subsection {Formula (77): An amazing breakthrough or a smoking gun? }

As you may recall, Formula (77) in the Google paper (Equation (\ref{e:77}), Section \ref {s:77})
provides an  estimation of the fidelity of a circuit based on the fidelities of its components:

$${\bf Formula~~(77)~~~}~~F~=~ \prod_{g \in {\cal G}_1} (1-e_g) \prod_{g \in {\cal G}_2} (1-e_g) \prod_{e \in {\cal Q}} (1-e_q).$$

The Google paper claims that this formula estimates with a precision of 10\%--20\% the
probability of the failure (fidelity) 
of a circuit. This remarkable agreement is a major new scientific discovery and
it is not needed for building quantum 
computers. Reaching sufficiently high fidelity levels is indeed crucial, but the demonstration of such  
accurate predictions on the fidelity based on the error rates of the individual components 
is neither plausible nor required. The precise fidelity estimation  
is only needed for the specific extrapolation argument leading to the Google
team's supremacy declarations.

In my opinion the claim regarding the fidelity estimation 
is 
very
implausible 
and  
even if quantum computers will eventually be built 
we are not going to witness the realization of this particular claim. 
Of course, it might be interesting to check whether we ever see anything remotely like this for 
other groups attempting to build quantum circuits, or indeed whether we ever see
in any other field of engineering 
such a good estimation of the failure probability of a physical system,
with hundreds of  interacting 
elements, as the product of hundreds of individual error probabilities.

The Google team's interpretation of this discovery is that it shows that 
there is ``no additional decoherence physics'' when the system scales,
and  they justify the remarkable predictive power of
their Formula (77) (Equation (\ref {e:77})) 
with a statistical computation that is based on the following three ingredients:

\begin {enumerate}
\item
Individual read-out and gate errors are accurate. 
The Google team reported that the level of accuracy for the individual qubit and gate fidelities is $\pm$20\%.
\item
Errors for the individual fidelity estimates are unbiased; namely, there are no systematic errors.
\item
  Error probabilities are statistically independent.\footnote {Based on these assumptions,
    Google's (rough) estimation 
 of the deviation of the prediction of Formula (77) is 
 \begin {equation}
   \label {e:fn}
 0.2 \cdot ( \sqrt n\cdot 0.038 + \sqrt{|{\cal G}_1|}\cdot 0.0016 + \sqrt{|{\cal G}_2|}\cdot 0.0063 ).
     \end {equation}
 (So, say, for $n=53$ and $m=14$ this gives roughly
   8.8\%.) However, The gap between the (77) prediction and the fidelity estimation based on the data,
   while bounded at 10\%-20\%, does not increase with $n$ as
   Formula (\ref {e:fn}) suggests.} 

         \end {enumerate} 

In my view all these claims are questionable
and the second and third claims are very implausible. This suggests that 
the excellent quality of the predictions based on Formula (77) may reflect naive statistical experimental 
expectations rather than physical reality.

{\bf A few remarks:} Let me first explain  the issue of biased versus unbiased 
estimation (the second item) with a simplified example. 
Suppose that you have a space rocket with 900 components and the probability of any component   
failing is estimated at 0.01. If one component fails, the entire space rocket fails.  
Under a statistical independence assumption, the probability 
of success is $(1-0.01)^{900}$, which roughly is 0.00012. If your estimate of 0.01 for 
each individual component is correct up to an unbiased error of 
20\% (namely, with probability 1/2 the correct error probability is 0.012 and 
with probability 1/2 it is 0.008),  
then the deviation of the outcome can be estimated within roughly 3\%.  
But if your estimation is systematically biased in one direction by 20\% then the effect on the 
probability of success is by a factor of five or so.

We also note that positive correlation between the error probabilities  
will actually lead to higher fidelity. There is, in fact, an entire discipline, 
in statistics and systems engineering, called reliability theory,
that studies failure properties of devices 
based on the failure distributions of individual components.

Finally, an explanation for the success of Formula (77), suggested by 
Peter Shor (in a discussion in my blog) and various other scholars \cite{Panel19}, is that 
the statistical independence needed for the success of Formula (77) 
is justified for random circuits. I do not see a justification 
for this claim, but it surely deserves further study.

\subsection {What needs to be done}

Listed below are steps required for a further assessment of the Google supremacy claims:
\begin {itemize}
\item
Further documentation of past experiments and a more careful documentation of future experiments. 

\item
  Replications of the experiments by the Google team: larger samples and further experiments
  in the classically tractable regime; further experiments in the 40--53 qubit range.

\item
Blind tests: some of the required replications by the Google team should apply the standard methodology of blind tests.

\item
Replications by other groups of various aspects of the Google claims, including the supremacy claims, 
the fidelity prediction claims, and the calibration methodology.\footnote{Here, I mean ``replications'' in
  a broad sense: replications by other groups need not apply the precise Google 2-qubit coupler.
  We can learn a lot from sampling based on a random circuit with standard 2-qubit gates,
  and if doing it for 53 qubits is too difficult, reliable experiments on 20--30 qubits could already be useful.
  A clear challenge would be to replicate (even in these easier settings) the
  prediction power of formula (77), or even something only ten times worse.}

\item
  Careful examination of the supremacy experiments both by the
  Google quantum-computing group itself, by the scientific community, and by Google. 
\end {itemize}

\subsection {Under the mathematical lens: Noise, variance, and Pythagoras}
\label{s:g2-ml}

Another aspect of the experiment that deserves thorough examination is 
the extent to which the noisy distributions presented by Google's experiment fit 
the theoretical expectation.
This is one aspect of the work I am currently conducting  with Yosi Rinott and Tomer Shoham \cite{RSK20}.
In this section we talk about several interesting mathematical and statistical aspects of  
distributions produced by NISQ circuits.

\subsubsection *{A toy model for the noise of quantum circuits}

Below is a simple toy model of what the noisy version of a quantum sampling problem may look like. 
It is based on the model
from Section \ref {s:aaqc-ml}.
Let $D(x_1,x_2,\dots, x_n)$ be a probability distribution on 0-1 vectors of length $n$.
Given a parameter $t$ we consider the noisy version of $D$ as

\begin {equation}
\label {e:nqc1}
N_t(D)(x)=\sum _{y \in \Omega_n}D(x+y)t^k(1-t)^{n-k}.
\end {equation}

Here, again, $y=(y_1,y_2,\dots,y_n)$ is also a 0-1 vector and $y_i=1$
indicates ``error in the $i$th coordinate.''
The sum $x+y$ should be considered as a sum modulo 2: $x_i+0=x_i$ and $x_i + 1= 1-x_i$.
If $E$ is a probability distribution on $\Omega_n$ then we can consider a more general form of noise, namely, 

\begin {equation}
\label {e:nqc2}
N_t(D)(x)=\sum _{y \in \Omega_n}D(x+y)E(y).
\end {equation}

Equation (\ref{e:nqc1}) is the case where $E(z)=B_t(z)=t^k(1-t)^{n-k},$ where $k=|z|$.
For random (or pseudorandom) quantum circuits, I expect that the effect of the noise on 
gates will be close to our model for the case where $E$
is a mixture of $B_t(y)$'s (more specifically, a Curie--Weiss distribution), 
and that this mixture will have a strong positive correlation between errors.
Modeling the noise by equations (\ref {e:nqc1}, \ref {e:nqc2}) abstracts away the dependence of noise
on the structure of the circuits and I expect that such modeling will be useful
both qualitatively and quantitatively.

\subsubsection *{The second-order term of noise}

Let us now move from an abstract study of noise to the Google experiment.
A simple approximation of the noisy distribution considered by Google is

\begin {equation}
\label {e:nn}
F D_C+ (1-F)U,
\end {equation}

\noindent
where $F$ is the fidelity. Namely, with probability $F$ we sample according to $D_C$ and with probability $(1-F)$ we sample according to the uniform 
probability distribution.

A more detailed description that we may expect is of the form 
\begin {equation}
F D_C+ (1-F) N_C,
\end {equation}

\noindent
where $N_C$ is a  small fluctuation of the uniform distribution that also depends on the circuit $C$. 
As it turns out, this more detailed form of noise does not affect 
Google's size-biased distribution and the $F_{EXB}$ estimator for the fidelity. 
Yet such detailed descriptions of the noise can be examined by performing similar tests 
specifically geared to the noise $N_C$.

Let us denote by $F_g$ the probability that no error occurs for 1-qubit or 2-qubit gates. 
We can split the noisy distribution into three parts,

\begin {equation}
\label {e:nr}
F D_C + (F_g-F)N_{RO} + (1-F_g)N_G,
\end {equation}

\noindent
where $N_G$ describes errors that involve also faulty gates, and $N_{RO}$ describes the
effect of read-out errors when there are no faulty gates. 
For the read-out errors, 
Equation (\ref {e:nqc1}) appears to give a good approximation, particularly
under Google's statistical independence assumption of read-out errors. 
Let $e_i$ denote the error probability for the $i$th qubit;
then,

\begin {equation}
\label {e:nr2}
(F_g-F)N_{RO} = (F_g-F) \sum_{y \in \Omega_n, y \ne 0} D_C(x+y)\prod_{i:y_i=1} (e_i) \prod _{i:y_i=0} (1-e_i).
\end {equation}

If we use averaged errors as in Equation (\ref{e:77s}) we reach a simpler formula.   
Let $F'= (1-0.0016)^{|{\cal G}_1|} (1-0.0063)^{|{\cal G}_2|} (1-0.036)^n$, 
and $F_g'= (1-0.0016)^{|{\cal G}_1|} (1-0.0063)^{|{\cal G}_2|}$. We replace $F'D_C+(1-F')U$ with 
$F'D_C+(F'_g-F')N'_{RO}+(1-F'_G)U$ with

\begin {equation}
\label {e:nr3}
(F'_g-F')N'_{RO} = (F'_g-F') \sum_ {y \in \Omega_n, y \ne 0} D_C(x+y)(1-0.036)^{|y|}(0.036)^{n-|y|}.
\end {equation}

\subsubsection *{Variance computation and Pythagoras }

Let me refer to a problem that was raised in relation to the
variance estimation of this statistical parameter. Given a circuit $C$
one can estimate the variance of the parameter for various samples.
However,  when considering the required size of samples for
several experiments for various circuits, one needs to compute the variance across different circuits,
while using the following formula: 

\begin {equation}
\label {e:pyt}
var (A)= \mathbb E (var (A|B)) + var (\mathbb E (A|B)). 
\end {equation}

When my friend and colleague Yosi Rinott teaches this formula for computing the variance, he tells the
students that they have surely seen this formula before.
For us it is an opportunity to see Greg Kuperberg's reference to 
Archimedes (Section \ref{s:g1-ml}) and raise him another 200 years (backwards) to
Pythagoras (c. 570 -- c. 495 BC).
Indeed, Equation (\ref {e:pyt}) is just a disguised form of the Pythagorean theorem.

\subsubsection *{A glimpse into my study with Yosi Rinott and Tomer Shoham}

1) Our study \cite{RSK20} of the fidelity estimate of the Google
team
confirms that a more precise
description of the noise (of the kind considered above) 
will not make a difference in the expected value
of $F_{XEB}$ and will make only a small insignificant difference in the 
variance. (Here the Pythagorean formula
for the variance (Equation (\ref {e:pyt})) comes into play.)
(In general, both $F_{XEB}$ and the entire size-biased empirical distribution are fairly robust.)
It also confirms and extends results
of the Google team asserting that compared to other (moment) estimators
of a similar nature, $F_{XEB}$ has smaller variance
and therefore smaller samples are required for definite results.

2) Google's samples  would allow
us to check on the data our proposal for the read-out noise, $N_{RO}$.
This provides an alternative fidelity estimator allowing to test the 
quality of the data of the Google experiment. (So far, this alternative fidelity
estimation has been checked only for $n=12,14$.)

3) A preliminary study of the Google data on 12 and 14 qubits further suggests that neither Google’s basic noise
model nor our refined read-out model fits the observed data, and
the second moment of the empirical distribution
is considerably higher than what the models predict.
On the other hand, there is a perfect agreement between
experiment and theory regarding the
size-biased distribution (Figure S32 in \cite {Arut+19S}) that also deserves examination.
For 12 and 14 qubits the data also exhibits non-stationary behavior (that might be chaotic). This seems consistent with
the noise-sensitivity pictures from Sections \ref {s:4f} and \ref {s:aaqc-ml} and deserves to be examined
for other NISQ samples.

4) Another finding from \cite {RSK20} that is also related
to the the analysis in \cite {Arut+19}(Section IV.A, especially Formulas (17,21)), is the following:
when one considers a probability distribution based on a
{\it specific} realization of a Porter--Thomas distribution, then the Google
statistic $F_{XEB}$ is no longer an unbiased estimator.
We asserted that when $x$ is drawn from the distribution $FD_C + (1-F)U$,
the expected value of $2^nD_C(x)$ is $1+F$ 
and, therefore, $F_{XEB}$ is an unbiased estimator for the fidelity $F$.
This assertion is correct over all realizations of the Porter--Thomas distribution
(or over all random circuits $C$), but for a specific realization (or a specific circuit $C$), $F_{XEB}$ is biased.
The expected value of $2^nD_C(x)$ is $1+\alpha F$, where
\begin {equation}
\alpha = -1 + 2^n \sum (D_C(x))^2.
\end {equation}
This leads to a similar yet better estimator (referred to as $V$) for the fidelity
that depends on the specific circuit $C$. 
In \cite {RSK20} we also study the maximum
likelihood estimator (MLE) which is superior to other estimators
mentioned here (and is also unbiased for every realization).
These observations suggest an interesting improvement of Google's main statistical
tool (when the number of qubits is not large).

\section {Possible connections and applications}
\label {s:app}

In this section we mention various potential applications and connections to physics arising
from a
fundamental failure of
quantum computation and quantum error-correction. 
Also here, the proposed connections and applications largely rely on the argument against quantum computers and a
fundamental failure of
quantum computation and quantum error-correction. Yet, a few of the insights described in this section can apply to 
fragments of quantum physics and quantum engineering even in the case where quantum computers are possible.
Finally, we explore also strange counterintuitive consequences of the reality without quantum computation that 
may even weaken the argument against quantum computers.

\subsection {Time and geometry}

For classical computers, the program you run is not restricted by the
geometry of the computer, and the information described by
a piece of your hard disc does not depend on the geometry of that piece.
This is such an obvious insight that 
we do not even spare it a second thought. Universal quantum computers will allow implementing  
quantum states and quantum evolutions on an array of qubits of arbitrary shape. 
On the other hand, the impossibility of quantum error-correction suggests that 
quantum states and evolutions constrain the geometry. The failure of quantum fault-tolerance 
will contradict computer-based intuitions  
that the information does not restrict the geometry, but will agree 
with insights from physics, where witnessing different 
geometries supporting the same physics is unusual and important. An example of
an important geometric distinction, 
when it comes to quantum behavior, is the different behavior of physics of different geometric scales: we witness very 
different microscopic physics, mesoscopic physics,
and macroscopic physics.

The same is true for time. With quantum fault-tolerance, 
every quantum evolution that can experimentally be created can be time-reversed and, 
in fact, we can permute the sequence of unitary operators describing the evolution in an arbitrary way. 
In a reality where quantum fault-tolerance is impossible, time reversal is not always possible

It is a familiar idea 
that since (noiseless) quantum systems are time-reversible, time emerges from quantum noise (decoherence).
(This idea has its early roots in classical thermodynamics.) 
Putting geometry and time together, we can propose that, generally speaking, 
quantum noise and the absence of quantum fault-tolerance 
enable the emergence of time and geometry.

\subsection {Superposition and teleportation}

In a recent paper about the future of physics, Frank Wilczek (2015) \cite {Wil15}  
predicts that large-scale quantum computers will eventually be 
built and describes why these excite him: 
``A quantum mind could experience a superposition 
of `mutually contradictory' states [...]  such a mind could revisit the past at will, and could be 
equipped to superpose past and present. To me, a more inspiring prospect than factoring large numbers.''

Indeed, superposition 
is at the heart of quantum physics --

and a common intuition that is supported by an ability to build universal quantum computers is that
for every two quantum states that can be constructed, their superposition can also be constructed.
Similarly, a common intuition is 
that every quantum state that can be prepared can also be teleported.

A central insight stemming from the argument against quantum computing (and the
various proposed laws associated with it)
is that already for a small number of qubits
certain pure states cannot be well approximated. (The fidelity $F$ is a good measure for
what ``well approximated'' means.)
For two pure states $\rho_1,\rho_2$ that can be achieved
but are close to the limit, a superposition between $\rho_1$ and $\rho_2$ that requires a more complicated
circuit than that needed for $\rho_1$ and $\rho_2$ may already be beyond reach.
By the same token, there is a quantum state $\rho$ that can be well approximated but is close
to the limit, and cannot be teleported. The reason is that
a circuit needed to demonstrate a teleportation for $\rho$ is considerably more involved
than a circuit needed to demonstrate $\rho$.

\subsection {Predictability and chaos}

Noise sensitivity asserts that for very general situations the effect of the noise will be devastating. 
This means that the actual outcomes not only will largely deviate from the
ideal (noiseless) outcomes but also will be very dependent on fine parameters of the noise, thus
leading to processes with large chaotic components.

\subsection {The black-hole information paradox}

Quantum information and computation play a role in explanations of the black-hole information
paradox.\footnote{In the absence of a definite theory of quantum gravity, the paradox can be seen as lying
between the foundation of physics and philosophy.}
Of particular importance in these explanations are ``pseudorandom'' quantum states
of the kind Google attempts to build
(but on a much larger number of qubits). According to our laws, such pseudorandom quantum states cannot be
achieved locally, and this goes against the rationale of some of the attempted solutions. On the other hand,
our laws asserting that A) qubits are inherently noisy and B)  entanglement
is necessarily accompanied by correlated noise 
may already suggest a resolution to some versions
of the ``paradox'' (e.g., to those based on no-cloning 
or on monogamy of entanglement).

\subsection {The time-energy uncertainty principle}

The time--energy uncertainty principle (TEUP) is a much-studied (controversial) issue in quantum mechanics.
Counterexamples 
were given by (Yakir) Aharonov and Bohm \cite {AhaBoh61}, and are based on the
ability to prescribe time-dependent quantum processes. 
A counterexample to an even weaker and more formal version of
TEUP was given by (Dorit) Aharonov and Atia \cite {AtiAha17} 
based on Shor's factoring algorithm.  Our study casts doubt on the very ability 
to prescribe noiseless time-dependent quantum evolutions at will,
while also challenging  the feasibility of
Shor's algorithm, and thus the picture drawn here in fact militates against the
physical relevance of these counterexamples.

\subsection {Realistic models for fluctuations}

One interesting property suggested by a critical look at the
theory of quantum fault-tolerance is that fluctuations in quantum systems
with an (even small) amount of interaction are super-Gaussian (perhaps even linear). Here, we
challenge one of the consequences
of the general Hamiltonian models allowing
quantum fault-tolerance (see, e.g., \cite {Pre13}). These models allow for some
noise correlation over time and space but they are
characterized by the fact that the error fluctuations are sub-Gaussian. Namely,
when there are $N$ qubits the standard deviation
for the number of qubit errors behaves like $\sqrt N$
and the probability of more than $t \sqrt N$ errors decays as it does for Gaussian distributions.

There are various quantum systems where the study of fluctuations will prove interesting.
For example, systems for
highly precise physical clocks are
characterized by having a  huge number $N$ of elements with extremely
weak interactions. We still expect (and this may even be supported
by current knowledge) that
in addition to $\sqrt N$-fluctuations there will also be
some $\epsilon N$-fluctuations. Of course, the relation between the
level of interaction and $\epsilon$ is of great interest. 
(The intuition of sub-Gaussian fluctuations may even be more remote from  
reality for engineering devices
and this is also related to our discussion of Google's Formula (77).)

\subsection {The unsharpness principle}

The unsharpness principle is a property of noisy quantum systems that
can be proved for certain quantizations of symplectic spaces.
This was studied by Polterovich (in \cite{Pol14}) who relies
on deep notions and results from symplectic geometry
and follows, on the quantum side,
some earlier works by Ozawa \cite {Oza04} and Busch, Heinonen, and Lahti \cite {BHL04}.
Here, 
the crucial distinction is
between general positive operator-valued measures (POVMs) and von-Neumann observables,
which are special cases of POVMs
(also known as projector-valued POVMs). The
unsharpness principle asserts that (under some locality condition) 
certain noisy quantum evolutions described by POVMs must
be unsharp, namely, ``far'' from von Neumann observables.
The amount of unsharpness is bounded from below by some non-commutativity measure.
It is interesting to explore the (mathematical and physical) scope of the unsharpness 
principle and its connection to our first law.

\subsection {Topological quantum computing}

Topological quantum computing is an approach whereby robust qubits are created not by
implementing quantum error-correction on NISQ circuits but by realizing stable qubits via anyons.
The argument from Section \ref{s:aaqc} can be extended 
also to this case (see \cite {Kal20}[Sec. 3.5]. In any case,
it is plausible that topological quantum computing and circuit-based
quantum computing will meet the same fate. (See also Section \ref {s:nd}.)

\subsection {Are neutrinos Majorana fermions?}

Majorana fermions are a type of fermions constructed mathematically by
Majorana in 1937 but so far not definitely detected in nature.
However, there is a compelling argument that neutrinos (or, more precisely,
an expected yet undiscovered heavy type of neutrinos) are Majorana fermions.

At the ICA workshop in Singapore, David Gross commented
that anyonic qubits required for topological quantum computing are based
on condensed-matter analogs of Majorana
fermions, which constitutes a strong argument that anyonic qubits are feasible.
Taking this analogy for granted, we can ask whether an argument against
topological quantum computing casts doubts on the
common (conjectural) expectations for Majorana fermions.
However, a review of the literature  (e.g., \cite {Bee19}) 
and consultations with colleagues revealed that Majorana fermions
from high-energy physics are most commonly regarded as
analogs of more mundane objects (Bogoliubov quasiparticles) from condensed-matter physics. 
Therefore, the argument against topological quantum computers and stable anyonic qubits
does not shed light on the nature of neutrinos
(but this is indeed the {\it kind} of insight we would {\it hope} to get).

\subsection {Noise stability and high-energy physics}

Extending the framework of noise stability and 
sensitivity to mathematical objects of high-energy physics is an appealing challenge.
Let us assume for a minute that this can be done.
We can ask if our second law asserting that realistic quantum states and evolutions 
are noise-stable provides some insights
into the various mysteries surrounding definite, but
unexplained, features of the standard model.

\subsection {Does nature support supersymmetry?}

Supersymmetry is a famous 
mathematical extension of the mathematics of the standard model.
It is widely believed that supersymmetry and, in particular, supersymmetric extensions of the standard model
are crucial to understanding 
physics beyond the standard model and quantum gravity.
So far, there is no definite experimental support for this belief.

Our second law imposes a severe limitation on quantum states and evolutions and asserts
that they can be described within a very restrictive computational class {\bf LDP} of low-degree polynomials.
We asked above whether this law can contribute to the understanding of the standard model, and we can ask
the same question with reference to the proposed supersymmetric extensions of the standard model.
Our second law supports classical error-correction and classical computation
but not quantum error-correction and quantum computation, and an appealing analogy might be that
the second law does not support supersymmetric extensions of the standard model at all.

\subsection {Cooling and exotic states of matter}

Noise stability, or the bounded-depth/low-degree polynomial description, 
may shed (pessimistic) light on the feasibility of 
various exotic states of matter.   
In some cases, such exotic states of matter are beyond reach, and, in other cases, 
the computational restriction may apply only to low-temperature states. 
(As the entropy increases, there are more opportunities to represent our state as a mixture of pure states 
that abide by the complexity requirement.) 
Within a symmetry class of quantum states (or for classes of states defined in a different way), 
noise stability, or the low-degree polynomial description, may
provide an absolute lower bound for cooling. 
An appealing formulation would be that for a class of quantum states 
the ``absolute zero'' temperature may depend on the class.

\subsection {The emergence of classical information and computation}
\label {s:maj}

It is an interesting question to find fundamental reasons for why quantum information
is more fragile than classical information, see \cite {Ter12}. We propose the following answer:
the class {\bf LDP} of functions and probability distributions that can be approximated by
low-degree polynomials does not support quantum advantage and quantum error-correction,
yet it still supports robust classical information, and
with it also classical communication and computation.
The ``majority'' Boolean function  
has excellent low-degree approximations and
allows for very robust classical bits
based on a large number of noisy bits (or qubits).
It is possible that every form of robust information, communication,
and computation in nature
is based on classical error-correction where
information is encoded by repetition (or simple variants of repetition)
and  decoded in turn by some variant of the  majority function.
(On top of this rudimentary form of classical error-correction,
we sometimes witness more sophisticated forms of classical error-correction.)

\begin{figure}
\centering
\includegraphics[scale=0.1]{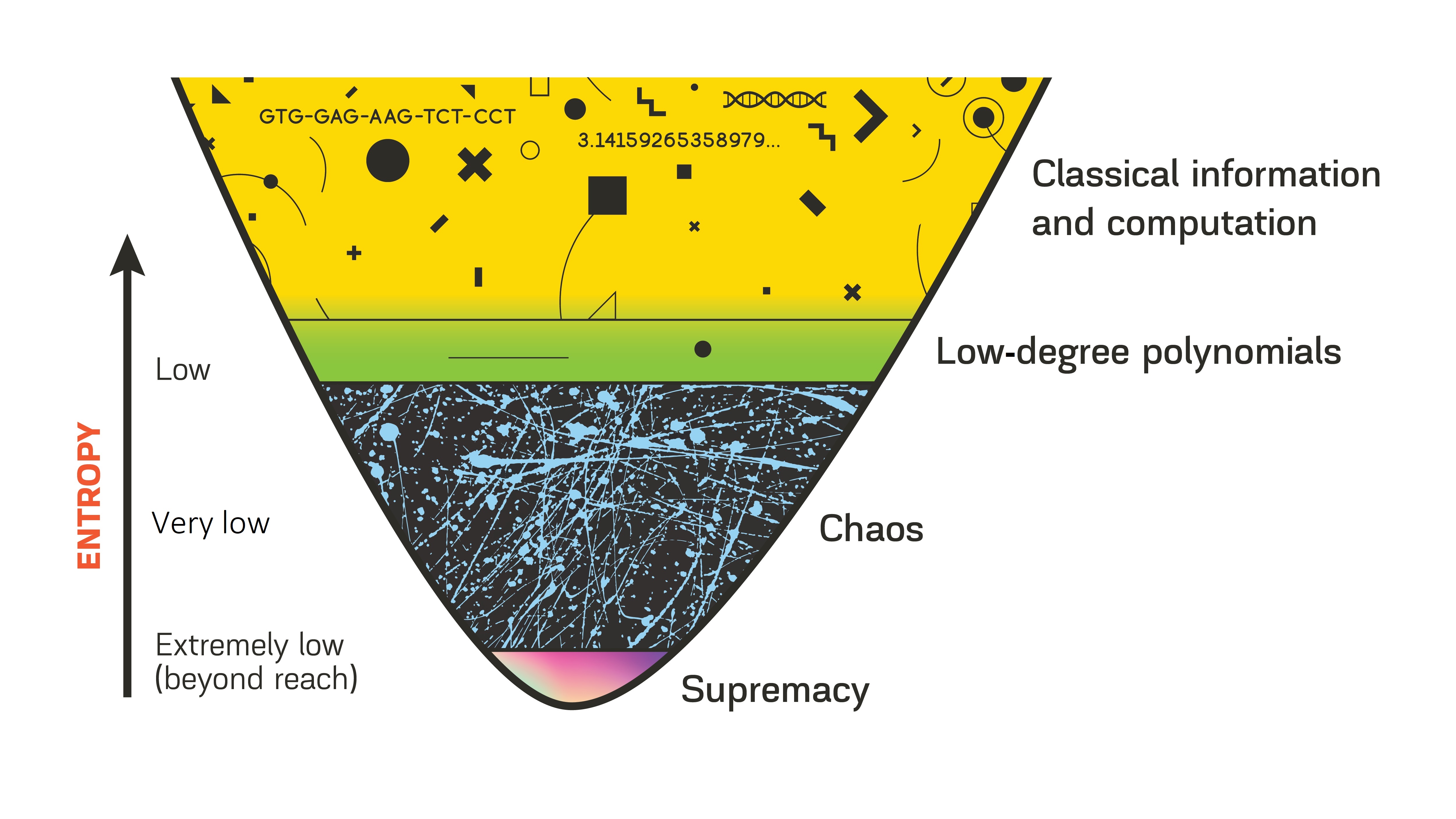}
\caption{Low-entropy quantum states give probability distributions described  by low degree polynomials, 
and very low-entropy quantum states give chaotic behavior. Higher entropy enables classical information.}
\label{fig:4}
\end{figure}

\subsection {Learnability of physical systems}

The theory of computing studies not only efficient computing but also efficient learning, namely, the 
ability to efficiently learn a member in a class from examples.
One major insight is that compared to carrying out computation when the model is known, it is
notably much harder to learn an unknown model. 
Efficient learning is very restrictive, but our very low-level class {\bf LDP} allows
for efficient learning. (We note that ${\bf AC^0}$
which is a
larger class and yet a very low level one does {\it not}
allow, in general, efficient learning. I don't know if there are approximately
learnable distributions beyond ${\bf AC^0}$. (See Fig. \ref{fig:1}.)
Efficient learnability of low-entropy quantum systems may provide an explanation for our ability to understand
natural processes and the parameters defining them.

A bold conjecture is that, in practice, robust distributions arising from NISQ systems are practically learnable via standard
machine-learning methods.

\subsection {Reaching ground states}

Reaching ground states is computationally hard ({\bf NP}-hard) for classical
systems, and even harder for quantum systems.
So how does nature reach ground states so often? 
Quantum evolutions and states
approximated by low-degree polynomials represent
severe computational restrictions that can make
reaching ground states computationally easy, and this
provides a theoretical support as to 
why, in many cases, nature easily reaches ground states.

\subsection {Noise and symmetry}

One insight from the failure of quantum error-correction and the accumulation of noise is that 
noisy quantum states and evolutions are subject to noise that respects their symmetries.

An interesting example is that of Bose--Einstein condensation. For a Bose--Einstein state on a bunch
of atoms, one type of noise corresponds to an independent noise for the individual atoms. 
Another type of noise represents fluctuations
of the collective Bose--Einstein state itself. This is the noise that respects the internal symmetries of
the state and it is expected that such a form of noise must always be present.

\subsection {Does Onsager's thermodynamic principle apply to quantum systems?}

(This connection was suggested by Robert Alicki years ago.)
Onsager's thermodynamical law expresses the idea that the 
statistical laws for the noise are related to the statistical laws for the ``signal.'' This 
idea is related to the effects of noise accumulation and to some of the items previously discussed. 
There is some controversy regarding the question of whether and how 
Onsager's law extends to quantum physics and it will be
interesting to see whether the proposed counterexamples are 
in conflict with our restrictions on noisy quantum processes. 

\subsection {The extended Church--Turing thesis}

The extended Church--Turing thesis (ECCT) (see, e.g., \cite {Wol85} and \cite {Pit90}) 
asserts that every realistic computing device can only 
perform efficient classical computation. Universal quantum computers 
violate the extended Church--Turing thesis.  By contrast, our theory
supports the validity of the extended Church--Turing thesis. 
See \cite {Kal20} for a detailed discussion. 
(We note that our theory is not based on the ECCT, but rather  
on computational complexity considerations for very low-level complexity classes.)

\subsection {Naturalness revisited}
\label {app-nat}
Here are three examples of similar deductions based on the
naturalness heuristic (Section \ref {s:nat1}) for computational complexity.

The first example is an important part of the theoretical foundation of the
Google experiment (See \cite {Arut+19,AarGun19}).

\begin {itemize}

\item [A1)]  
  Finding a sample with $F_{XEB} > \epsilon$ is exponentially hard as a function
  of $n$ (for a fixed $\epsilon$).

\item [A2)]  
This supports the assertion that achieving this task (for $\epsilon =1/1000$) on 53 qubits
represents quantum advantage.  

\end {itemize}

The second example refers to a recent proposal for
implementing Shor's factoring algorithm using classical devices called
stochastic magnetic circuits \cite {BPF+19}.

\begin {itemize}

\item [B1)]
The computational power of the stochastic magnetic circuits
offered for implementing Shor's algorithm  is within {\bf P}.

\item [B2)]
This supports the assertion that these devices offer no superior way to factor integers.

\end {itemize}

And, finally, the third example
is the crux of my argument against quantum computers.

\begin {itemize}
\item [C1)]
The computational power of NISQ computers is {\bf P} (for a fixed rate, $\epsilon$, of noise).

\item [C2)]  
This supports the assertion that NISQ computers offer no superior computation.

\end {itemize}

The naturalness heuristic plays (often in an implicit way) a central role in the way computational complexity insights are related to
computational reality. It is relevant to computational complexity insights in practical algorithms, in scientific computing,
in practical areas of cryptography, and in machine learning and statistics. This is an interesting topic for further study.

\subsection {``So what about the energy levels of the lithium atom?''}

The argument against superior quantum computation suggests that 
robust computations performed by nature can, at least in principle, be carried out efficiently on a
digital computer. Yet, there are robust physical quantities that ``nature computes''
for which efficient classical computations (and especially computations ``from first principles'')
are currently unavailable.
(For more on this issue, see \cite {Kal16b}[Sec. 6.5 and Sec. 4.] or \cite {Kal16}.)

\subsection {Correlation and modeling the noise}
\label{s:n-cor}

A critique of the third law reads as follows:

``Entanglement is a feature of a state (in Hilbert space), not of the operator that acts on the state.
The noise is due to which operator acts on the state.
In quantum error-correction and fault-tolerance theory we
analyze the structure of the operator that acts on the state and show that the locality
of the interactions in this operator and the weakness of the unwanted interactions enable fault-tolerance,
not whether it [the operator] acts on entangled states or product states. Locality of interactions here means
that we have no 10-body interactions, etc.: really every Hamiltonian, field theory, or
theory that is ever used in physics is in accordance with this notion,
so deviating from this concept seems ill-advised and badly motivated.''

This point deserves to be explained: nothing in the theory described here is based on non-local modeling of the noise.
As a matter of fact, it is based on the very standard modeling of noisy quantum circuits.
Our argument (Section \ref{s:aaqc}) asserts that $\beta < \delta$ and therefore quantum error-correction is not possible.
Now, it is well accepted both as part of the theory and as an empirical fact that 
when we create entanglement for two qubits directly by a gate we face correlated errors:
depolarizing noise that collapses the state to the maximal entropy state for the four-dimensional
Hilbert space describing the pair of qubits. What the third law simply says is that in the
absence of quantum error-correction 
the accumulated errors will be correlated (provided they are still small enough)
also for entanglement created indirectly.

\subsection {``It from qubit'': Does entanglement explain geometry and gravity?}
\label {s:ste} 

Over the past decade, there have been several proposals (often referred to as ``it from qubit'')  
that gravity (and other parts of physics) can be understood from 
insights and techniques derived from quantum information theory and particularly entanglement. 
People have raised questions like:
Does spacetime emerge from entanglement? Can entanglement shed light on gravity? And
can quantum computers simulate all physical phenomena?

The idea that spacetime emerges from entanglement is in line with the concept that  
quantum states restrict time and geometry. 
Yet, the type of entanglement presented in some of these works 
is often well beyond the reach of local quantum processes according to our viewpoint.
Some proposed connections between spacetime and entanglement might be consistent with a (speculative) 
possibility that nature
is described by more than one  local system 
when certain states that are mundane for one local system are highly 
entangled for other systems. 

\subsection {Theory, reality, and practice}

Many of the items listed in this section may lead to interesting mathematics, and I hope to put some of them 
under the mathematical lens or, better yet, to see this done by others.  
Let me suggest a wider context for the discussion, one that encompasses  
understanding the relation between 
theory, reality, and practice in computer science, in physics,
and in other applications of mathematics.\footnote {The relations 
between the theory of computing and practical reality was one of the themes in my ICM2018 paper \cite {Kal18}, 
and it is based on three examples: linear programming, voting methods, and quantum computers.}

\section {Developments in recent months (written: March 2021)}
\label {s:nd}
\subsection {The photonic advantage claim}

  A recent paper \cite {Zho+20} published in {\it Science} claims to achieve ``quantum computational advantage''
  at room-temperature using photons. Specifically,
  the paper reports a  Gaussian boson sampling experiment representing a quantum state
  in $\sim 10^{30}$-dimensional Hilbert space 
  and a sampling rate that is $\sim 10^{14}$ faster than that of using digital supercomputers.
  This paper was described as the first independent verification of the Google's quantum advantage claims. In fact, the
  claimed advantage is several orders of magnitude higher than Google's claims.
  
  This huge computational advantage claim is based on
  certain statistical tests measuring the proximity of the empirical samples to
  the outcomes of noiseless
  simulations of the quantum experiment. (The simulations were run on a digital supercomputer.)
  The statistical reasoning of \cite {Zho+20}) 
  is
  based on comparing the empirical samples
  to a few other distributions.
  However, in view of the 2014 results of Kalai and Kindler \cite{KalKin14} this statistical reasoning is incorrect and therefore
  the conclusion of achieving huge quantum computational advantage is unfounded.
  Moreover, a polynomial-time
  algorithm from Kalai and Kindler \cite {KalKin14} may achieve similar or better
  sampling quality for the statistical methods of \cite {Zho+20}. See also \cite {KalKin21} and Renema \cite {Ren20} (and papers cited there).
  Kalai and Kindler's analysis \cite {KalKin14,KalKin21} is based on taking a truncated
  Fourier--Hermite expansion on the boson sampling distribution. Renema's paper proposes another  method of ``spoofing,'' namely,
  another efficient algorithm for achieving similar sampling quality based on an algorithm of Clifford and Clifford \cite {CliCli17}.

  \subsection {Quantum advantage via quantum annealing?}

  Recent claims \cite{KRL+21}
  by scientists from D-Wave and Google have asserted that quantum annealing algorithms performed by a D-Wave quantum computer
  on several problems are several orders of magnitude faster compared to certain classical packages for the same problems.
  These claims add to earlier claims by D-Wave scientists from 2014 and 2018. 
  Since, in this case, there is no theoretical foundation for quantum advantage, D-Wave
  claims have been received with skepticism by the quantum computing community.

  \subsection {Are stable non-abelian anyons possible?}

  Topological quantum computing is based on creating protected qubits via anyons. As we already mentioned,
  my argument can be extended to imply that protected topological qubits are not possible either (\cite {Kal20} Section 3.5).
  Now, an important step toward creating protected topological qubits is the creation of
  certain condensed-matter quasi-particles, and this was claimed in a 2018 paper \cite {Zha+18} by
  researchers from Microsoft.
  While my argument directly contradicts the huge computational advantage claims,
  the situation here is more nuanced: I do not know if
  the quantum states claimed in \cite {Zha+18} can lead to sampling that demonstrates a quantum advantage, or are
  in conflict with some other laws proposed here.
  This question may depend on the level of noise, and certainly deserves further study. Be that as it may be,
  the authors of \cite {Zha+18} have retracted \cite {Zha+21} their claims
  because of inconsistencies between the raw measurement data and the figures that were published in the paper.

  \subsection {``Spoofing'' Sycamore}

  Pan and Zhang \cite {PanZha21} proposed a general tensor network method for simulating quantum circuits.
  As an application, they studied the sampling problem of Google's Sycamore circuits, and announced that
  by using a moderate computing power
  they could generate one million (very) correlated bitstrings from the Sycamore circuit with 53 qubits and 20 cycles,
  with (XEB) fidelity equal to 0.739, which is much higher than the fidelity in Google's quantum supremacy experiments.   
  This result 
  on its own may shed serious doubts on Google's supremacy claims.

\subsection {Under the mathematical lens: Sampling and matching}
  
All four items described above are related to fascinating mathematics. I will mention only one simple
connection that fits in nicely with several topics discussed in this paper. Consider  a bipartite graph with a set $A$ of $n$ vertices
on one side and a set $B$ of $m$ vertices on the other side, $m \ge n$.
Suppose that you want to sample a multi-subset $C$ of $B$ according to the number $n(A,C)$ of semi-matchings from $A$ to $C$.
Here, a multi-subset is
a list of elements from $B$ with repetitions (see, e.g., \cite {HLLT06}).
If $C$ is a multi-subset of $B$, a semi-matching is a map from $A$ to $C$
such that every vertex is mapped to a neighbor, and the images
are precisely the vertices in $C$ with the prescribed multiplicities.
When $C$ is an ordinary subset this is the ordinary notion of matching (see Section \ref{s:match}). 
Now, computing $n(A,C)$ is computationally hard ({\bf \#P-complete}).
However, there is a simple efficient algorithm for the sampling task. (The sampling is on the nose.) The algorithm is based on \cite {CliCli17}
and I learned it from Jelmer Renema. Simply, choose for each vertex of $A$ a neighbor in $B$ uniformly at random!   
This small note gave us an opportunity to consider again matchings that
are most fascinating mathematical objects, with great importance in theoretical computer science,
and to be reminded that sampling is easier (and can be much easier in some cases) than computing the probabilities.
(The computational gap between sampling and
computing individual probabilities was already a main insight of Troyansky and Tishby \cite {TroTis96}.)

\section {Conclusion}

The crux of the argument against quantum computation is simple.
For fixed constant error rates, quantum circuits in the intermediate
scale are primitive computational devices. They represent computation in {\bf P} and, more
than that, a computational complexity class {\bf LDP}
that even allows polynomial-time learnability.
This implies that a huge quantum computational advantage is beyond reach for NISQ computers, and therefore the harder task
of creating good-quality quantum error-correcting codes is beyond reach as well.
This argument is robust since for a
large range of sub-constant error rates we can expect
chaotic (noise sensitive) behavior.
The argument presented here explains why classical computation is possible:
the computational class {\bf LDP} supports rudimentary forms of classical error-corrections, and therefore, in the large scale,
also robust classical information and
computation. The argument applies to all the different proposals for implementing NISQ computers and
extends to forms of quantum
computation that are not based on NISQ devices such as topological quantum computing.
My argument predicts that the
recent claims of a huge quantum computational advantage are false.
Preliminary work by others and myself supports this prediction.

My work on quantum computation started in 2005 and is marked by three major stages. Until 2013 I mainly
studied correlations of errors (for entangled states) and my efforts
could be described (in hindsight) as mainly
trying to draw conclusions from the failure of quantum fault-tolerance. Some of
those conclusions are described
in Sections \ref {s:laws} and \ref {s:app}.
The connection to noise stability and noise sensitivity, leading to my computational theoretic argument
against quantum computers
arose from my 2014 work with Guy Kindler on boson sampling.
Conducting a large part of the discussion in English,
while at times placing some fragments under the mathematical lens,
is characteristic not only of this paper but of my work as a whole.

As of the end of 2019, my argument against quantum computers was challenged by
a bold far-reaching experimental claim.
Seeking to critically study and possibly refute the Google claims is
different from merely seeking to understand the laws of abstract noisy quantum systems.
Having an opportunity to
rethink matters of statistics (with my colleagues Yosi Rinott, Tomer Shoham, and others)
is pleasant, but, on 
the other hand, 
trying to understand what is really going on in the Google experiment is also, in various
ways, less uplifting.
Yet, I also find this pursuit to be of interest and importance
that extends 
beyond the specific case in question.
I wish to stress that my critique of the Google experiment was first brought to the attention
of the Google team and discussed with them.
In the skepticism and debate that have swirled around quantum computing and that I have been
involved with in
the past 15 years, 
winning has not been the only thing; indeed, it has not even been 
the most important thing. 
What 
I find important is making the right choices and right judgments in
delicate scientific and social situations that are full of uncertainties.

Over the past four decades, the very idea of quantum computation
has led to many advances in several areas of physics, engineering, computer science,
and mathematics. I expect that the most important application will eventually be the
understanding of the impossibility of quantum error-correction and quantum computation.
Overall, the debate over quantum computing is a fascinating one,
and I can see a clear silver lining: major advances in human ability to
simulate quantum physics and quantum chemistry are expected to emerge
if quantum computational advantage can be demonstrated and quantum computers can be built,
but also 
if quantum computational advantage
cannot be demonstrated and quantum computers cannot be built.

Some of the insights and methods characteristic of the area of quantum
computation
might be useful for 
classical computation of realistic quantum systems -- which is, apparently, what nature does.


\begin{thebibliography}{99}




\bibitem 
{AarArk13} S. Aaronson and A. Arkhipov,
The computational complexity of linear optics, {\it Theory of Computing} 4 (2013), 143--252.

\bibitem {AarGun19} S. Aaronson and S. Gunn,
  On the classical hardness of spoofing linear cross-entropy benchmarking, (2019),  arXiv:1910.12085. 




\bibitem {AhaBen97} D. Aharonov and M. Ben-Or,
Fault-tolerant quantum computation with constant error, in STOC '97,
ACM, New York, 1999, pp. 176--188.


\bibitem {AhaBoh61} 
Y. Aharonov and D. Bohm, 
Time in the quantum theory and the uncertainty relation for time and
energy, {\it Physical Review,} 122 (1961), 1649--1658.


\bibitem {AGK19} 
R. Arratia, L. Goldstein and F. Kochman, Size bias for one and all,  
{\it Probability Surveys} 16 (2019), 1--61. arXiv:1308.2729.

\bibitem {Arut+19} 
F. Arute et al., Quantum supremacy using a programmable superconducting processor, 
{\it Nature,} 574 (2019), 505--510.

\bibitem {Arut+19S}
  F. Arute et al. (2019),
Supplementary information for ``Quantum supremacy using a programmable 
superconducting processor'' (2019) arXiv:1910.11333. 

\bibitem {AtiAha17} 
Y. Atia and D. Aharonov, 
Fast-forwarding of Hamiltonians and exponentially precise measurements,
{\it Nature Communications,} (2017) Article Number 1572, arXiv:1610.09619.

\bibitem {Bee19} C. W. J. Beenakker, Search for non-Abelian Majorana
  braiding statistics in superconductors, arXiv:1907.06497.



\bibitem{BKS99}
I.~Benjamini, G.~Kalai, and O.~Schramm,
\newblock Noise sensitivity of Boolean functions and applications to
percolation,
\newblock {\em Publications Math\'ematiques de l'Institut des Hautes \'Etudes
Scientifiques} 90 (1999), 5--43.

\bibitem {BPF+19}
  W. A. Borders, A. Z. Pervaiz, S. Fukami, S. et al.,
  Integer factorization using stochastic magnetic tunnel junctions, {\it Nature} 573 (2019), 390--393. 


  
\bibitem {BHL04} P. Busch, T. Heinonen, and P. Lahti, Noise and disturbance in quantum
measurement, {\it Physics Letters A} 320 (2004), 261--270.


\bibitem {CliCli17} P. Clifford and R. Clifford, The classical complexity of boson sampling
(2017) arXiv:1706.01260.



\bibitem {Fey82} R. P. Feynman, Simulating physics with computers,
{\it International Journal of Theoretical Physics} 21 (1982), 467--488.

\bibitem {HLLT06} N. Harvey, R. Ladner, L. Lov\'asz, and T. Tamir,  Semi-matchings for bipartite graphs and load balancing,
  {\it Journal of
Algorithms} 59 (2006), 53--78.

\bibitem {Hua+20}
  C. Huang et al.,
  Classical simulation of quantum supremacy circuits (2020), arXiv:2005.06787.


\bibitem {Panel19} S. Irani (Moderator), Supremacy Panel, 
Hebrew University of Jerusalem, Dec. 2019.
Participants: D. Aharonov, B. Barak, A. Bouland, G. Kalai,
S. Aaronson, S. Boixo, and U. Vazirani.
https://youtu.be/\_Yb7uIGBynU .



\bibitem {KKL88}
J. Kahn, G. Kalai, and N. Linial, The influence of variables
on Boolean functions, in {\it Proceedings of the 29th Annual Symposium on Foundations of
Computer Science}, 1988, pp. 68--80.



\bibitem {Kal16} G. Kalai, The quantum computer puzzle, {\it Notices
of the American Mathematical Society}
63 (2016), 508--516.

\bibitem {Kal16b} G. Kalai, The quantum computer puzzle (expanded version), arXiv:1605.00992.

\bibitem {Kal18} G. Kalai, Three puzzles on mathematics, computation and
games, in {\it Proceedings of the International
Congress of Mathematicians} 2018, Rio de Janeiro, Vol. I, 2018, pp. 551--606.

\bibitem {Kal20}
  G. Kalai, The argument against quantum computers, in: M. Hemmo, and O. Shenker,(eds.) 
{\it Quantum, Probability, Logic: Itamar Pitowsky's Work and Influence}, 
Springer(2020),pp. 399--422, arXiv:1908.02499.

\bibitem{KalKin14} G. Kalai and G. Kindler, Gaussian noise sensitivity
and BosonSampling (2014), arXiv:1409.3093.

\bibitem {KalKin21} G. Kalai and G. Kindler, Concerns about recent claims of a huge quantum computational advantage via Gaussian
  boson sampling, preprint 2021.

\bibitem {KRL+21}
  A. D. King, J. Raymond,T.  Lanting, et al.
  Scaling advantage over path-integral Monte Carlo in quantum simulation of
  geometrically frustrated magnets. {\it Nature Communications} 12 (2021),
  Article Number 1113. 

\bibitem {Kin75} J. F. C. Kingman, Random discrete distributions,
  {\it Journal of the Royal Statistical Society,
Series B} 37 (1975), 1--22.

\bibitem{Kit97} A.~Y.~Kitaev, Quantum error
correction with imperfect gates, in {\it Quantum Communication,
Computing, and Measurement
}, Plenum Press, New York, 1997, pp. 181--188.


\bibitem{KLZ98}
E.~Knill, R.~Laflamme, and W.~H.~Zurek, Resilient
quantum computation: Error models and thresholds, {\it Proceedings
of the Royal Society of London A }{454} (1998), 365--384.


\bibitem {Kup19} G. Kuperberg, Archimedes’ other principle and quantum supremacy, 
Guest post on ``Shtetl Optimized,'' Nov. 2019. 


\bibitem {Oza04} M. Ozawa, Uncertainty relations for joint measurements of noncommuting
observables, {\it Physics Letters A} 320 (2004), 367--374.

\bibitem {PanZha21} F. Pan and P. Zhang, 
Simulating the Sycamore quantum supremacy circuits (2021), arXiv:2103.03074.



\bibitem{Pit90}
I. Pitowsky, The physical Church thesis
and physical computational complexity, {\it lyuun, A Jerusalem
Philosophical Quarterly} 39 (1990), 81--99.

\bibitem {PGN+19}
E. Pednault, J. A. Gunnels, G. Nannicini, L. Horesh, and R. Wisnieff, 
Leveraging secondary storage to simulate deep 54-qubit Sycamore circuits (2019), 
arXiv:1910.09534.


\bibitem {Pol14} L. Polterovich,  Symplectic geometry of quantum
noise, {\it Communications in Mathematical Physics} 327 (2014), 481--519.


\bibitem {PorTho56}
C. E. Porter and R. G. Thomas, 
Fluctuations of nuclear reaction widths, 
{\it Physical Reviews} 104 (1956), 483--491.

\bibitem{Pre13} J. Preskill,
Sufficient condition on noise correlations for scalable quantum computing,
{\it Quantum Information and Computing} 13 (2013), 181--194.

\bibitem {Ren20} J. Renema, Marginal probabilities in boson samplers with arbitrary input states (2020), arXiv:2012.14917.


\bibitem {RSK20} Y. Rinott, T. Shoham, and G. Kalai, Statistical aspects of the quantum supremacy
demonstration (2020), arXiv:2008.05177.



\bibitem {Sho94} P. W. Shor, Polynomial-time
algorithms for prime factorization and
discrete logarithms on a quantum computer,
{\it SIAM Review} 41 (1999), 303--332.
(Earlier version,
{\it Proceedings of the 35th Annual Symposium on Foundations of
Computer Science}, 1994.)

\bibitem {Sho95} P. W. Shor, Scheme for reducing
decoherence in quantum computer
memory, {\it Physical Review A} 52 (1995), 2493--2496.


\bibitem {Ste96}
A.~M.~Steane, Error-correcting codes in
quantum theory, {\it Physical Review Letters} 77 (1996), 793--797.


\bibitem {Ter12} B. Terhal, The fragility of quantum information?
 In: A. H. Dediu, C. Mart\'in-Vide, B. Truthe (eds),
{\it Theory and Practice of Natural Computing. TPNC 2012.}
  Lecture Notes in Computer Science, Vol. 7505. Springer, Berlin, 2012, pp. 47--56, arXiv:1305.4004.

\bibitem {TroTis96} L. Troyansky and N. Tishby, Permanent uncertainty: On the quantum evaluation
of the determinant and the permanent of a matrix, in {\it Proceedings of the 4th Workshop 
on Physics and Computation}, 1996.



\bibitem {Wil15} F. Wilczek, Physics in 100 years, arXiv:1503.07735 (2015).  

\bibitem {Wol85} S. Wolfram, Undecidability and intractability
in theoretical physics, {\it Physical Review Letters} 54 (1985), 735--738.

\bibitem {Zha+18}
  H. Zhang, C.X. Liu, S. Gazibegovic, et al.,
Quantized Majorana conductance, {\it Nature} 556 (2018), 74--79. 


\bibitem {Zha+21}
  H. Zhang, C.X. Liu, S. Gazibegovic, et al.,
  Retraction Note: Quantized Majorana conductance, {\it Nature} (2021).

\bibitem {Zho+20}
  H.-S. Zhong, H. Wang, Y.-H. Deng et al., Quantum computational advantage using photons, {\it Science} 370 (2020) 1460--1463.

\bibitem {ZSW20}
Y. Zhou, E. M. Stoudenmire, X. Waintal, 
What limits the simulation of quantum computers? (2020), arXiv:2002.07730.



\end{thebibliography}
\end {document}